%% file: vtcfall_2019.tex
\documentclass[conference, a4paper]{IEEEtran}
\input{macro}

\usepackage[caption=false,font=footnotesize]{subfig}

\IEEEoverridecommandlockouts

\begin{document}

%
% paper title
\title{Enabling Ultra Reliable Wireless Communications for Factory Automation with Distributed MIMO}
\author{
\IEEEauthorblockN{Gianluca Casciano\IEEEauthorrefmark{1}, Paolo Baracca\IEEEauthorrefmark{2}, and Stefano Buzzi\IEEEauthorrefmark{1}}
\IEEEauthorblockA{\IEEEauthorrefmark{1}University of Cassino and Southern Lazio, Cassino, Italy\\ 
\IEEEauthorrefmark{2}Nokia Bell Labs, Stuttgart, Germany}
}

\maketitle
\begin{abstract}
Factory automation is one of the most challenging use cases for 5G-and-beyond mobile networks due to strict latency, availability and reliability constraints. In this work, an indoor factory scenario is considered, and distributed multiple-input multiple-output (MIMO) schemes are investigated in order to enable reliable communication to the actuators (ACs) active in the factory.
Different levels of coordination among the access points serving the ACs and several beamforming schemes are considered and analyzed. To enforce system reliability, a max-min power allocation (MPA) algorithm is proposed, aimed at improving the signal to interference plus noise ratio (SINR) of the ACs with the worst channel conditions. Extensive system simulations are performed in a realistic scenario, which includes a new path-loss model based on recent measurements in factory scenarios, and, also, the presence of non-Gaussian impulsive noise. Numerical results show that distributed MIMO schemes with zero-forcing (ZF) beamforming and MPA have the potential of providing SINR gains in the order of tens of dB with respect to a centralized MIMO deployment, as well as that the impulsive noise can strongly degrade the system performance and thus requires specific detection and mitigation techniques.
\end{abstract}

\section{Introduction}
\label{sect_intro}

Machine-type communications for factory automation are realized nowadays mainly through wired technologies \cite{wollschlaeger_iem17}, but the recent past has witnessed a growing interest towards their replacement with wireless communications for several reasons.
Indeed, the installation and maintenance of cables is much more expensive than the deployment of a wireless network. Additionally, the adoption of  wireless communications allows much higher flexibility in the production deployment inside the factory, enabling, for instance, the deployment of swarm of robots autonomously moving around.
Wireless factory automation has thus become one of the most important use cases for the fifth generation (5G) and beyond mobile networks \cite{holfeld_cm16}.
Differently from  previous wireless network generations, that were targeting mainly high data rate requirements for mobile broadband traffic, 5G has been indeed designed from its start to cope with mixed traffic types, each one characterized by specific requirements in terms of rate, latency, availability, and reliability.
In fact, the term ultra reliable and low latency communications (URLLC) refers to all the use cases that pose very strict latency and reliability requirements.
The third generation partnership project (3GPP) has already identified several scenarios for factory automation, with motion control being the one with the most extreme requirements for the latency (down to $1$ ms) and availability/reliability (up to $99.9999\%$) \cite{3gpp_ts22261}.
First attempts to introduce wireless communications for factories have been done in unlicensed frequencies because of the huge chunks of spectrum available.
On the other hand, solutions in licensed bands represent a better option to meet these very strict performance requirements, since in unlicensed frequencies there are usually severe constraints on the maximum transmit power and, also, uncontrolled interference from other networks.

The capabilities of a 5G radio interface for factory automation have been first analyzed in \cite{yilmaz_icc15}, showing both the high potentials of 5G and the necessity of some interference mitigation mechanisms.
Coverage and capacity have been analyzed then for different deployment strategies in \cite{brahmi_globecom15}, where a simple frequency planning is shown as already very beneficial in improving the performance.
Even better reliability can be obtained when applying power control \cite{singh_pimrc16a} and multi-hop relaying (even with just a maximum of two hops) \cite{singh_pimrc16b}.

Overall, in order to increase reliability, the signal to interference plus noise ratio (SINR) at the receiver must be increased, and diversity techniques appear to be the most suited strategy to achieve this goal \cite{johansson_icc15}.
On the other hand, time and frequency diversity represent solutions with non-negligible limitations for URLLC: time diversity, being based on re-transmissions, has marginal applicability in use cases like motion control with strict latency constraints, whereas frequency diversity is limited by the available system bandwidth. Spatial (antenna) diversity instead appears to be an effective strategy to meet the reliability requirements of URLLC.

In this work, we consider a factory scenario where a central controller coordinates a set of access points (APs) to communicate with the actuators (ACs) active in the factory. While previous works in this area have considered basic interference mitigation mechanisms like frequency planning \cite{brahmi_globecom15} and power control \cite{singh_pimrc16a}, we assume that the APs form a distributed multiple-input multiple-output (MIMO) system, with the central controller coordinating their transmission and reception. Focusing on the communication link from the APs to the ACs -- the customarily defined downlink -- different deployments of APs with varying number of antennas are evaluated. For these deployments, several beamforming schemes are considered and evaluated under the assumption of single AP transmission (SAT) and of joint transmission (JT). 
A max-min power allocation (MPA) method aimed at maximizing the minimum SINR across the ACs is also considered. The performance study of the several proposed APs deployments, beamforming schemes, and of the MPA rule is carried out in a realistic 5G scenario, exploiting a recently proposed path-loss model for industrial environments, based on measurements done at 3.5 GHz in operational factories \cite{tdoc_1813177}. This represents a sharp contrast with previous works \cite{yilmaz_icc15, brahmi_globecom15, singh_pimrc16a, singh_pimrc16b}, that have used the channel model in \cite{tanghe_twc08}, based on measurements taken in unlicensed bands.
The path-loss model of  \cite{tdoc_1813177}, instead, perfectly fits the $3.5-3.7$ GHz band that is widely recognized as the main candidate for factory automation with 5G \cite{qualcomm_whitepaper}. In addition, the effect of non-Gaussian impulsive noise, which can severely limit the performance in the industrial environment \cite{sanchez_isie07}, is also included in our study.

The obtained numerical results will show that distributed MIMO schemes with MPA and JT can strongly improve the system availability, whereas the impulsive noise can be a source of performance degradation that requires specific detection and mitigation techniques to be properly handled.

{\em Notation}: We use $\left( \cdot \right)^{T}$ and $\left( \cdot \right)^{H}$ to denote transpose  and conjugate transpose, respectively. $\bm{I}_{N}$ denotes the identity matrix of size $N$, $\bm{0}_{N\times M}$ the matrix of size $N\times M$ with all zero entries, $\left\| \bm{X} \right\|$ the Frobenius norm of matrix $\bm{X}$, $\left[\bm{X}\right]_{n,m}$ the entry on row $n$ and column $m$ of $\bm{X}$, $\left[\bm{X}\right]_{\cdot,m}$ the $m$th column of $\bm{X}$, and $\left[\bm{x}\right]_{n}$ the $n$-th entry of vector $\bm{x}$.

\section{System Setup}
\label{sect_system}

We assume a typical industrial scenario similar to the one in \cite{singh_pimrc16a} with a factory hall of dimension $100 \times 50 \times 6\,\mbox{m}^3$. The system has a total available bandwidth of $80$ MHz at a carrier frequency of $3.5$ GHz, but for simplicity we simulate just a sub-band of $10$ MHz, so as to  leverage the flat-fading approximation. We denote by $K$ the number of single antenna ACs uniformly dropped in the factory hall at a height of $2$ m.

\begin{figure}[!t]
\centering
\subfloat[$J=1$.]{\includegraphics[width=0.7\hsize]{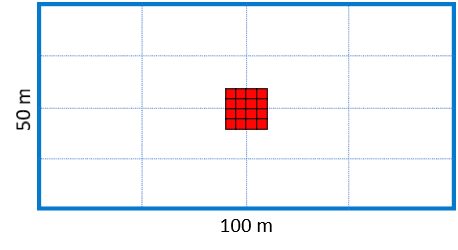}
\label{fig_j1}}
\hfil
\subfloat[$J=4$.]{\includegraphics[width=0.7\hsize]{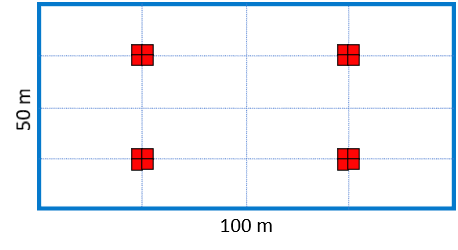}
\label{fig_j4}}
\hfil
\subfloat[$J=16$.]{\includegraphics[width=0.7\hsize]{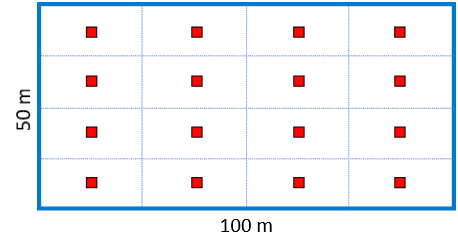}
\label{fig_j16}}
\caption{Compared AP deployments, with a red square representing 4 antennas.}
\label{AP_deployment}
\end{figure}

We assume to have  $M_{\rm TOT}=64$  antennas that can be organized in different APs deployments (see also Fig. \ref{AP_deployment}). In particular, denoting by $J$ the number of APs and by $M$ the number of antennas at each AP, with $M=M_{\rm TOT}/J$, the following architectures are considered.
\begin{itemize}
\item[a)] A {\it centralized} deployment with $J=1$ AP equipped with $M=64$ antennas and located in the center of the factory hall. In this case the factory can be seen as a single-cell network served by one base station with a co-located large antenna array.
\item[b)] A {\it partially distributed} deployment with $J=4$ APs, each equipped with $M=16$ antennas.
\item[c)] A {\it fully distributed} deployment with $J=16$ APs, each equipped with $M=4$ antennas.
In this and in the previous case b), the factory can be seen as a multi-cell wireless network.
\end{itemize}
The height of the APs is 6 m and the position of the APs with $J = 4, 16,$ is determined by dividing the factory hall in $J$ rectangular coverage areas, all with the same shape and orientation.

As already discussed, we focus on  downlink transmission (i.e., AP-to-AC link) and assume that the  total transmit power is $30$ dBm \cite{yilmaz_icc15}, which results in $\left[ P_{\rm AP} \right]_{\rm dBm} = 21$ dBm on the simulated sub-band. 
Since the total transmit power is independent of the number $J$ of deployed APs, the comparison among the above three cases a), b) and c) is fair both in terms of total transmit power and total number of transmit antennas. 
%However, a distributed deployment allows anyhow to typically achieve a better coverage as the AP-AC distance is smaller when compared to a centralized deployment, thus reducing the path loss.

The noise power $\sigma_{k}^2$ at the $k$-th AC can be written as
\begin{equation}
\sigma_{k}^2 = \sigma_w^2 + \sigma_{k,i}^2\,,
\end{equation}
where $\sigma_w^2$ is power of the customary thermal noise, which is modeled as a white Gaussian process with power spectral density of $-174$ dBm/Hz, while $\sigma_{k,i}^2$ is the power of the impulsive noise (more details about it in Section \ref{subsect_impnoise}). Moreover, an additional receiver noise figure of 7 dB is considered.

The channel model is implemented based on the findings of \cite{tdoc_1813177}: in that paper the results of several measurements taken at $3.5$ GHz in two types of operational factories and with different AP deployments are detailed. These measurements are then used to upgrade the 3GPP indoor office model and provide more specific parameters for the path-loss, shadowing and line-of-sight (LOS) probability functions. Here, among the scenarios listed in \cite[Tab. 3]{tdoc_1813177}, we focus on the ``dense factory" with ``clutter-embedded APs" setup, which is characterized by the most harsh channel conditions.
Regarding the small scale fading,  it is worth noting that, as observed in \cite{tdoc_1813177, tanghe_twc08}, in  industrial environments there is an abundance of highly reflective metallic  materials, which introduce strong multi-path fading components. Accordingly, we assume a Rayleigh small scale fading model, i.e., the $1 \times M$ channel vector between the $k$-th  AC  and the $j$-th AP  can be written as $\bm{h}_{k,j} \sim \mathcal{CN} \left( \bm{0}_{1 \times M}, \beta_{k,j}\bm{I}_M \right)$, with $\beta_{k,j}$ being the large scale fading attenuation generated according to \cite{tdoc_1813177}.

\subsection{Transmission Modes and Beamforming Techniques}
\label{subsect_tm}

We consider two different transmission modes (or AC-AP association rules) to serve the ACs.

As a baseline, we assume {\it single AP transmission (SAT)}, where  AC $k$ is served by its anchor AP $j_k$, which is the AP experiencing the strongest link toward that AC, i.e.,
\begin{equation}
j_k = \arg\max_{j=1,2,\ldots,J} \beta_{k,j}\,.
\end{equation}
Note that with SAT the backhaul infrastructure connecting the APs is required to provide the data to be sent to AC $k$ just to its anchor AP $j_k$. With SAT,  the considered system is similar to a wireless cellular network wherein each base station takes care of the mobile devices in its coverage area and with no coordination among the base stations. 

A possible alternative to SAT is {\it joint transmission (JT)} among all the APs deployed in the factory hall. 
In this case, each AC is served by \textit{all} the APs in the system. 
Differently from SAT, with JT the backhaul needs to provide the data to be sent to AC $k$ to all the APs: in fact, JT poses more stringent requirements in terms of latency and capacity on the backhaul infrastructure. With JT, the considered system is similar to a wireless network with advanced base station coordination strategies.
Of course, when considering the centralized deployment, the SAT and JT modes end up being coincident.

Besides the transmission modes, an additional level of interference mitigation can be realized through proper beamforming schemes to be used at the APs. In this work, we compare three beamforming criteria that have been widely used in the MIMO literature \cite{comp_tutorial}. 
For the sake of brevity, we provide the expressions of the beamforming vectors for the JT mode, since the extension to SAT mode can be done in a straightforward way. 

Let us thus denote first by $\bm{h}_k = \left[ \bm{h}_{k,1},\bm{h}_{k,2},\ldots,\bm{h}_{k,J} \right]$ the vector of size $1 \times M_{\rm TOT}$ collecting the channels between AC $k$ and all the APs deployed in the factory hall.

The first beamformer that we consider is the  {\it maximum ratio transmission (MRT)} one. In this case, the beamformer used to serve AC $k$ is designed to maximize the signal to noise ratio, neglecting interference, and can be written as
\begin{equation}
\bm{g}_k^{\rm (MRT)} = \frac{ \bm{h}_k^H }{ \left\| \bm{h}_k \right\| }\,.
\end{equation}

Another possible beamforming design strategy is the \textit{zero forcing (ZF)} one, that aims at nulling interference at the expense of a reduction of the useful signal margin against additive noise. After defining $\bm{H} = \left[ \bm{h}_1^T, \bm{h}_2^T, \ldots, \bm{h}_k^T \right]^T$ and $\bm{G} = \bm{H}^H \left( \bm{H} \bm{H}^H \right)^{-1}$, the ZF beamformer used to serve AC $k$ can be shown to be written as
\begin{equation}
\bm{g}_k^{\rm (ZF)} = \frac{ \bm{G}_{\cdot,k} }{ \left\| \bm{G}_{\cdot,k} \right\| }\,.
\end{equation}

When considering SAT mode, each AP forces to zero the interference caused to the ACs that are under its coverage area, neglecting the interference caused to the other ACs in the system. This may cause some performance degradation and leads to the third beamforming strategy that we consider, i.e., the so-called \textit{coordinated zero forcing} (CZF), which turns out to be useful in the SAT mode. In CZF, each APs creates nulls not only toward the served ACs, but also toward the other ACs in the factory hall. Notice that for JT mode the ZF and CZF beamformers coincide since all the ACs are served by the same set of APs, that act as a single AP with distributed antennas.

By denoting with $P_k$ the power allocated to AC $k$ and with $\mathbf{g}_k$ the beamformer intended for transmitting to AC $k$, the SINR for AC $k$ with JT can be easily shown to be written as
\begin{equation}
{\rm SINR}_k^{\rm (JT)} = \frac{\left| \bm{h}_k \bm{g}_k \right|^2 P_k }{ \sigma_{k}^2 + \sum_{m=1,m\neq k}^K \left|\bm{h}_k \bm{g}_m \right|^2 P_m }\,.
\label{eq_sinr}
\end{equation}

\subsection{Imperfect Channel State Information Model}

So far, perfect knowledge of the channel vectors has been assumed. 
In real systems, however, $\bm{h}_{k,j}$ is not known at the APs and must be estimated, with both SAT and JT, in order to design the beamformer. The way the channel state information (CSI) is obtained at the APs strongly depends on the duplexing mode: in frequency division duplex (FDD) the channels are estimated at the AC, quantized and then sent back to the APs, whereas in time division duplex (TDD) the channels are directly estimated at the APs thanks to training sequences sent by the ACs. While FDD is favorable for URLLC, since it allows a quicker access to the channel, TDD in general allows a better CSI knowledge at the APs. If the sequence of uplink and downlink slots in the TDD frame is designed in a way that allows meeting the latency constraint, TDD is a favorable solution as well for URLLC since it avoids channel quantization by the ACs.

Accordingly, we focus on the TDD protocol, and assume that the $K$ ACs are assigned orthogonal training sequences, so as to avoid pilot contamination \cite{galati_wcnc18} in the uplink channel estimation process.

Letting $\left[ P_{\rm AC} \right]_{\rm dBm} = 20$ dBm \cite{yilmaz_icc15} denote the AC transmit power during training, and assuming that the pilot sequences have a (discrete) duration  $T \geq K$,  the minimum mean square error (MMSE) estimate of $\bm{h}_{k,j}$, 
say $\hat{\bm{h}}_{k,j}$, can be written as \cite{wild_vtcfall11}
\begin{equation}
\hat{\bm{h}}_{k,j} = \frac{\gamma_{k,j}T}{1+\gamma_{k,j}T} \left( \bm{h}_{k,j} + \bm{z}_{k,j} \right) \,,
\label{eq_impcsi}
\end{equation}
where $\gamma_{k,j} = P_{\rm AC} \beta_{k,j} / \sigma_{{\rm AP},j}^2$ and $\bm{z}_{k,j} \sim \mathcal{CN} \left( \bm{0}_{1 \times M}, \sigma_{{\rm AP},j}^2 / \left( P_{\rm AC} T \right) \bm{I}_M \right)$, with $\sigma_{{\rm AP},j}^2$ being the noise power at AP $j$.

After estimating the channels, beamforming design is thus performed at the APs by replacing in Section \ref{subsect_tm} $\bm{h}_{k,j}$ and $\bm{h}_{k}$ with $\hat{\bm{h}}_{k,j}$ and $\hat{\bm{h}}_{k} = \left[ \hat{\bm{h}}_{k,1}, \hat{\bm{h}}_{k,2}, \ldots, \hat{\bm{h}}_{k,J} \right]$, respectively.

\subsection{Impulsive Noise Model}
\label{subsect_impnoise}

Electrical and other types of machines commonly used in industrial environments have been recognized as a source of heavy interference, in addition to the customary white Gaussian thermal noise, that can severely affect the communication at radio frequency bands \cite{sanchez_isie07}. This type of interference is typically modeled as a non-Gaussian impulsive noise with specific properties in terms of bandwidth, repetion and spacing of these pulses. Several models have been proposed to describe such non-Gaussian noise \cite{bhatti_eusipco09}; in this paper, we will  use the basic model proposed in \cite[Sect. V]{wang_cl97}, where the power $\sigma_{k,i}^2$ of the impulsive noise at AC $k$ is a random variable that can be written as
\begin{equation}
\sigma_{k,i}^2 = \Gamma \sigma_w^2 B_k \left( \epsilon \right)\,,
\label{eq_impnoise}
\end{equation}
where $\Gamma$ denotes the ratio between the impulsive noise and thermal white noise powers, $\epsilon$ is the probability that an impulsive noise event occurs during the transmission, and $B_k \left( \epsilon \right)$ is a Bernoulli random variable with mean $\epsilon$. Note that this model assumes the impulsive noise to be wideband and with a constant power over the system band. Although very basic, the tuning of the parameters $\Gamma$ and $\epsilon$ in (\ref{eq_impnoise}) allows having a first understanding of the impact of the impulsive noise on the performance achieved by the different distributed MIMO schemes.

\section{Max-min Power Allocation}
\label{sect_mpa}

We now focus on the power allocation strategy. Indeed,  after the beamforming strategy has been chosen, in (\ref{eq_sinr}) we still have the opportunity to optimize the power $P_k$ allocated to each AC $k$.
A basic criterion commonly used in cellular networks is called equal power allocation (EPA) and simply assumes the available power to be evenly split among the active ACs, i.e.,
\begin{equation}
P_k^{\rm (EPA)} = \frac{ P_{\rm AP} }{ K }\,.
\label{eq_epa}
\end{equation}
While this solution is widely used, it has strong limitations with URLLC, since, given the reliability and service availability constraints, it is crucial  to improve the tail of the SINR distribution \cite{bennis_proc18} rather than its average or median value. 
%Therefore, when looking at the performance of a single AC, the system must improve the SINR of the transmissions with the worst channel conditions, even by strongly sacrificing the performance in the slots where the AC experiences better channel conditions. 
Accordingly,  the system should focus on  improving the performance of the ACs in the worst conditions even by sacrificing the performance of the ACs in the best conditions, as long as they anyhow manage to meet the reliability constraints. It seems thus natural to consider the max-min power allocation (MPA) criterion, where the powers in the JT case (the extension to SAT is basic and not reported here for the sake of brevity) are the solution to the following optimization problem:
\begin{subequations}
\begin{equation}
\max_{P_k,k=1,2,\ldots,K} \min \frac{\left| \hat{\bm{h}}_k \bm{g}_k \right|^2 P_k }{ \sigma_{k}^2 + \sum_{m=1,m\neq k}^K \left|\hat{\bm{h}}_k \bm{g}_m \right|^2 P_m } \,,
\label{fo_mpa}
\end{equation}
$\qquad\qquad\qquad$ s.t.
\begin{equation}
\sum_{k=1}^{K} P_k \leq P_{\rm AP}\,,
\label{co1_mpa}
\end{equation}
\begin{equation}
P_k \geq 0\,,\quad k=1,2,\ldots,K\,.
\label{co2_mpa}
\end{equation}
\label{optimization_mpa}
\end{subequations}
Note that in problem (\ref{optimization_mpa}) the objective function (\ref{fo_mpa}) maximizes the minimum estimated SINR among the active ACs, whereas the constraint (\ref{co1_mpa}) imposes the transmit power to not exceed the maximum value $P_{\rm AP}$.
This is a well-known problem in optimization theory and its solution can be derived in our setup following the procedure defined in \cite[Sect. 3]{yang_icassp98}. In particular, upon defining the $K \times K$ matrix with entries
\begin{equation}
\left[ \bm{R} \right]_{k,m} = \begin{cases} \| \hat{\bm{h}}_k \bm{g}_m \|^2 / \| \hat{\bm{h}}_k \bm{g}_k \|^2, & \mbox{if } k \neq m \\ 0, & \mbox{if } k = m \end{cases}\,,
\end{equation}
and the two $K \times 1$ vectors $\bm{f} = \left[ \sigma_{1}^2 / \|\hat{\bm{h}}_1 \bm{g}_1\|^2, \ldots, \sigma_{K}^2 / \|\hat{\bm{h}}_K\bm{g}_K\|^2 \right]^T$ and $\bm{q} = \left[ \|\bm{g}_1\|^2, \ldots, \|\bm{g}_K\|^2\right]^T$, the solution can be obtained by first constructing the matrix
\begin{equation}
\bm{D} = \begin{bmatrix} \bm{R} & \bm{f} \\ \bm{q}^T\bm{R}/P_{\rm AP} & \bm{q}^T\bm{f}/P_{\rm AP} \end{bmatrix}\, ,
\end{equation}
and, then, denoting by $\bm{w}$ the eigenvector of the matrix $\bm{D}$ corresponding to the eigenvalue with the largest norm, the power allocated to AC $k$ with MPA turns out to be
\begin{equation}
P_k^{\rm (MPA)} = \frac{ \left[\bm{w}\right]_k }{ \left[\bm{w}\right]_{K+1} }\,.
\end{equation}

\section{Performance Evaluation}
\label{sect_results}

\begin{figure}[t!]
\centering
\includegraphics[width=1\hsize]{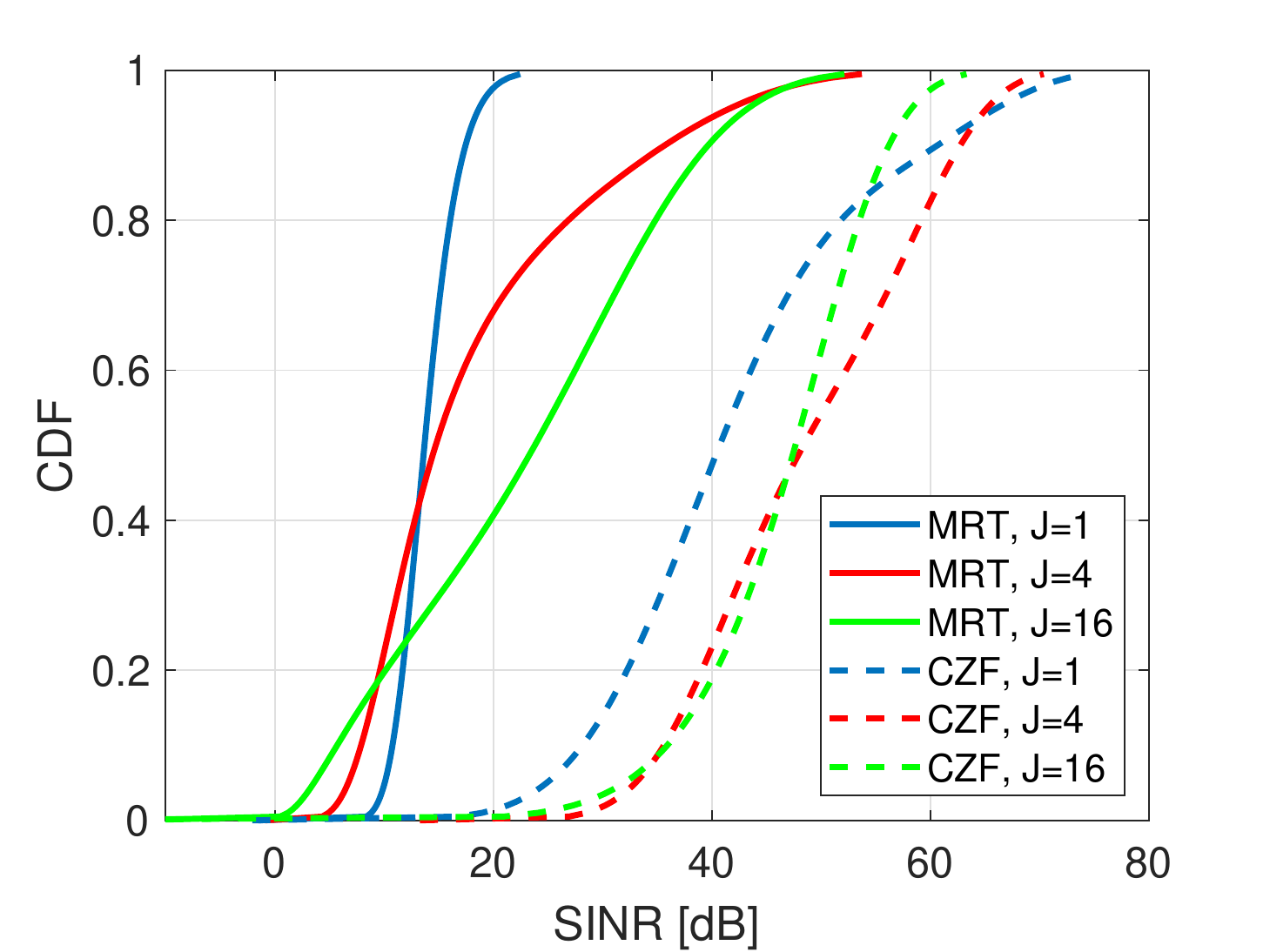}\\
\caption{CDF of the SINR with SAT for different deployments and beamformers by assuming PCSI, EPA, and $K=4$.}
\label{cdf_sinr_sct_epa_pcsi_k4}
\end{figure}

\begin{figure}[t!]
\centering
\includegraphics[width=1\hsize]{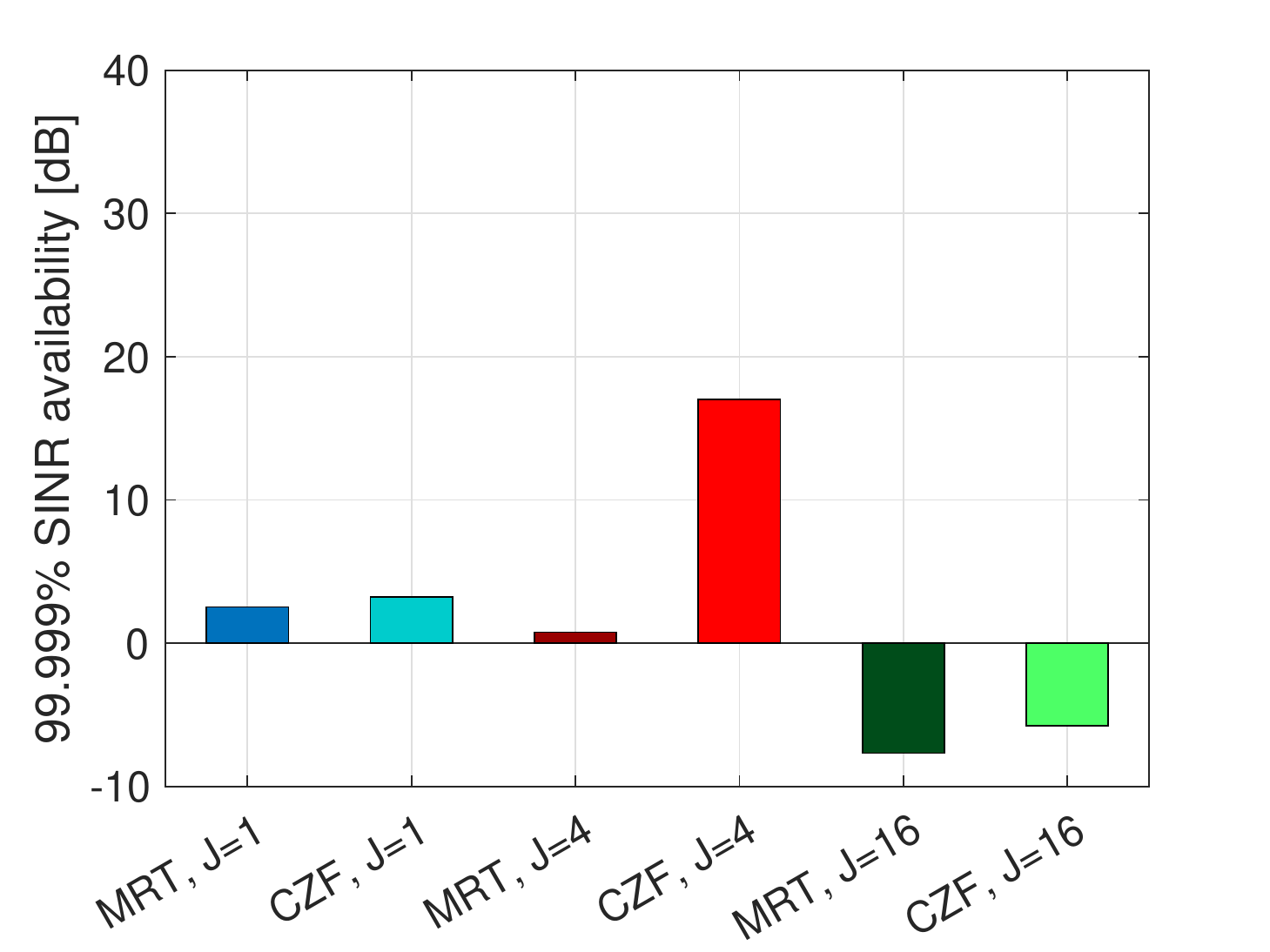}\\
\caption{SINR availability with SAT for different deployments and beamformers by assuming PCSI, EPA, and $K=4$.}
\label{sinr_availability_bf_sct_epa_pcsi_k4}
\end{figure}

In Fig.s \ref{cdf_sinr_sct_epa_pcsi_k4} and \ref{sinr_availability_bf_sct_epa_pcsi_k4} we compare the performance achieved by the different AP deployments when considering SAT, $K=4$ ACs served with EPA and perfect CSI (PCSI) knowledge at the APs. While in Fig. \ref{cdf_sinr_sct_epa_pcsi_k4} we report the cumulative distribution function (CDF) of the SINR, in Fig. \ref{sinr_availability_bf_sct_epa_pcsi_k4} we highlight the maximum SINR value that is achieved with $99.999\%$ probability, i.e., the $99.999\%$ SINR availability, which corresponds to the first $100\,000$-quantile of the CDF in Fig. \ref{cdf_sinr_sct_epa_pcsi_k4}. Note that, together with the block error rate (BLER) for the reliability, this is the most meaningful key performance indicator (KPI) when looking at system availability for URLLC: in the following, for the sake of space, we refer to this KPI just as {\it SINR availability}. First, we observe that CZF outperforms MRT by canceling the interference, with a gain up to about 16 dB with $J=4$ in the SINR availability. Then, the results highlight the tradeoff that needs to be considered when deploying the APs. More APs, i.e., a higher $J$, means a smaller AC-AP distance, which reduces the path-loss, but at the same time, with SAT, provides less beamforming gain and diversity (as each AP is equipped with just $M=M_{\rm TOT}/J$ antennas). On the contrary, a smaller $J$ introduces more path-loss but also more beamforming gain and diversity. Then, understanding which deployment is the best depends also on the beamforming criterion that is implemented. Indeed, by considering for instance CZF and comparing $J=1$ and $J=4$ in Fig. \ref{cdf_sinr_sct_epa_pcsi_k4}, we observe that $J=4$ provides the best performance for most of the ACs, and in particular for the ones with the worst channel conditions, whereas $J=1$ provides better SINR just for the ACs in the best conditions. On the other hand, when considering MRT, the best deployment for the ACs in worst conditions is obtained with $J=1$. Finally, we would like to highlight the very bad SINR achieved with CZF and $J=16$ in Fig. \ref{sinr_availability_bf_sct_epa_pcsi_k4}. Although in this setup we have $M=K=4$ and CSI is perfectly known at the APs (and indeed the SINR is, from Fig. \ref{cdf_sinr_sct_epa_pcsi_k4}, more than $20$ dB for most of the ACs), it might happen that two channels are almost parallel, which makes the computation of the inverse for CZF and ZF an issue. Although this event is rare, the SINR availability exactly captures these unlikely phenomena, which are however fundamental in understanding the performance with URLLC.

\begin{figure}[t!]
\centering
\includegraphics[width=1\hsize]{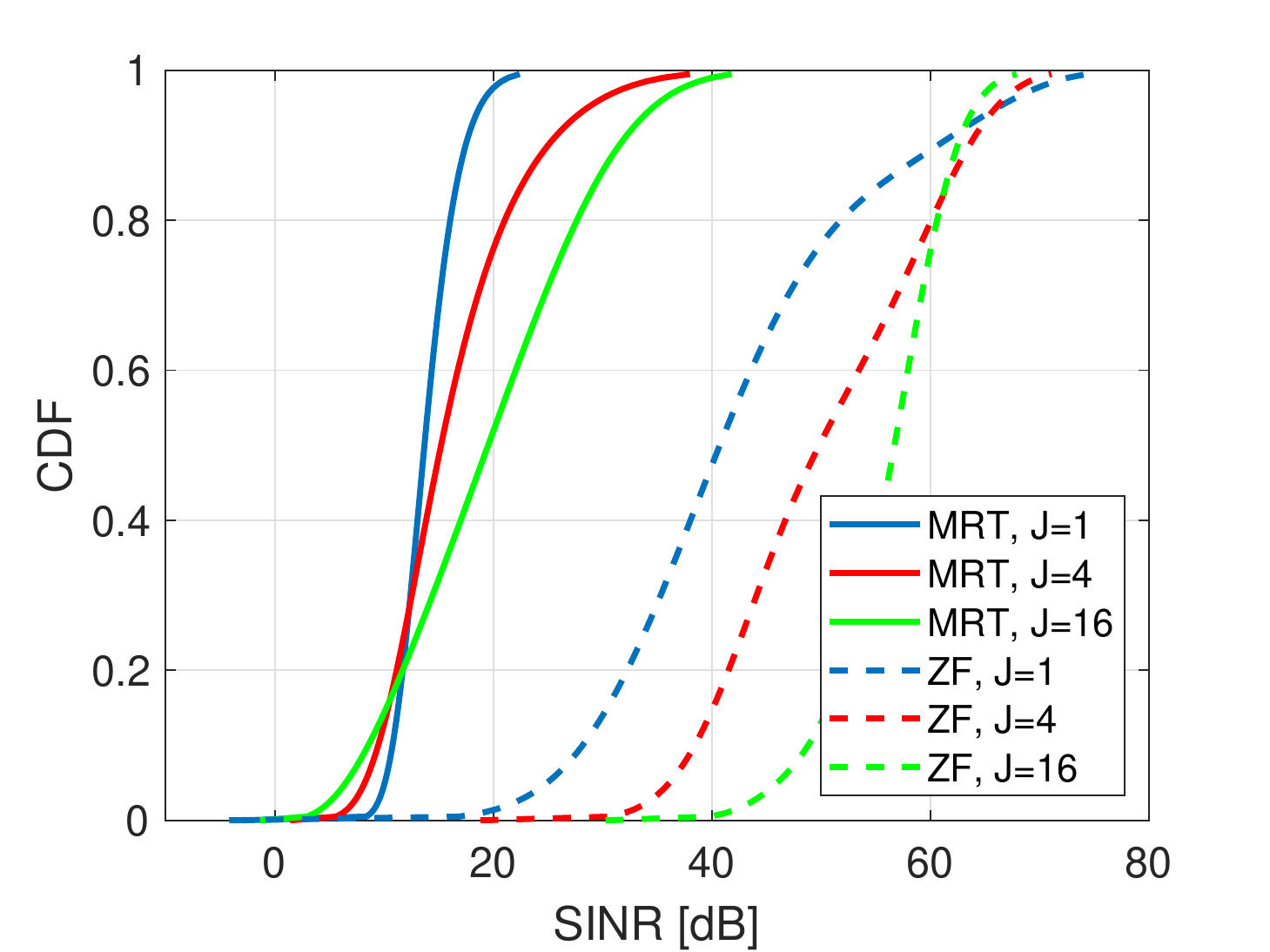}\\
\caption{CDF of the SINR with JT for different deployments and beamformers by assuming PCSI, EPA, and $K=4$.}
\label{cdf_sinr_jt_epa_pcsi_k4}
\end{figure}

\begin{figure}[t!]
\centering
\includegraphics[width=1\hsize]{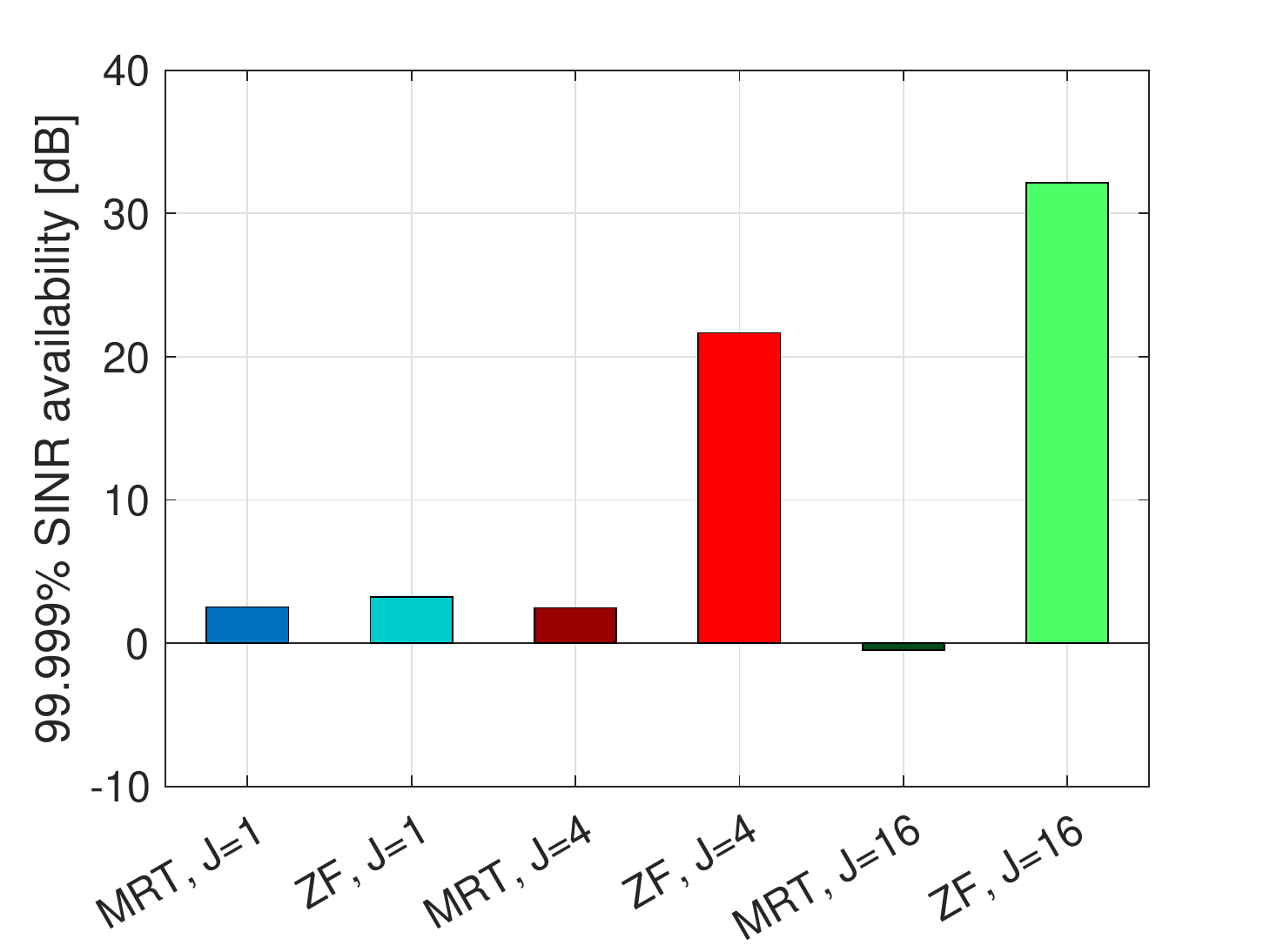}\\
\caption{SINR availability with JT for different deployments and beamformers by assuming PCSI, EPA, and $K=4$.}
\label{sinr_availability_bf_jt_epa_pcsi_k4}
\end{figure}

In Fig.s \ref{cdf_sinr_jt_epa_pcsi_k4} and \ref{sinr_availability_bf_jt_epa_pcsi_k4} we consider JT and compare different deployments with EPA, PCSI and $K=4$. Also with JT, ZF strongly outperforms MRT, showing the importance of interference mitigation mechanisms. Similarly to SAT, $J=1$ is the best deployment when using MRT. On the other hand, with ZF distributed MIMO deployments strongly outperform centralized MIMO by providing a gain of about $19$ and $29$ dB with $J=4$ and $J=16$, respectively.

Fig.s \ref{cdf_sinr_sct_epa_pcsi_k4}--\ref{sinr_availability_bf_jt_epa_pcsi_k4} show that, when interference is properly mitigated by the beamformer, the best deployment strategy is to reduce the path-loss by moving the APs closer to the ACs with distributed MIMO. On the other hand, when MRT is used and interference is not properly canceled, having a central AP with many antennas provides more fairness among the ACs and, as a consequence, the best SINR availability: that happens mainly because of a stronger uncontrolled interference affecting distributed MIMO deployments.

\begin{figure}[t!]
\centering
\includegraphics[width=1\hsize]{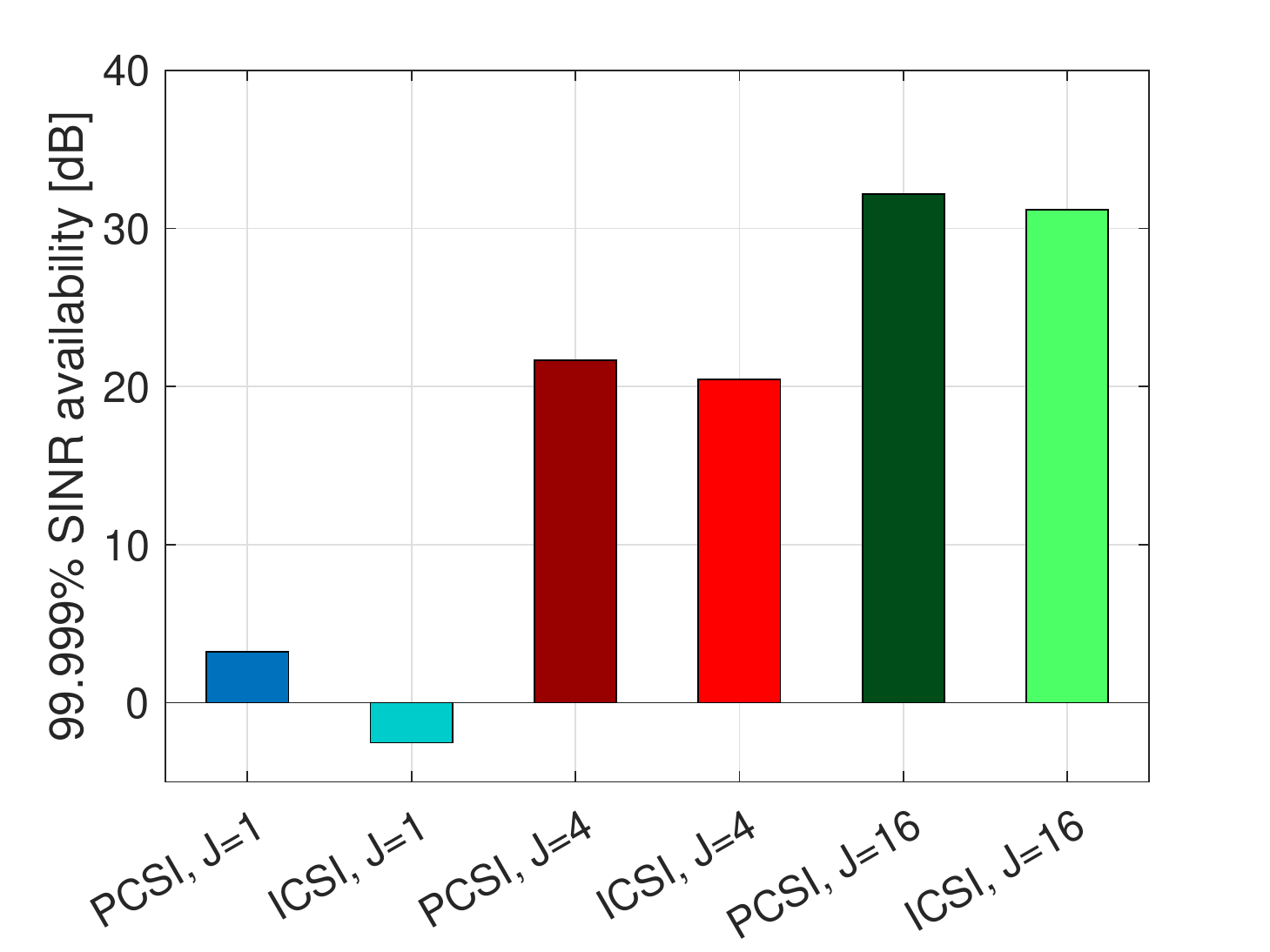}\\
\caption{Impact of ICSI on the SINR availability with JT and ZF for different deployments by assuming EPA and $K=T=4$.}
\label{sinr_availability_csi_jt_zf_epa_k4_t4}
\end{figure}

For the same setup of Fig.s \ref{cdf_sinr_jt_epa_pcsi_k4} and \ref{sinr_availability_bf_jt_epa_pcsi_k4}, we assume $T=K$ in (\ref{eq_impcsi}) and evaluate the impact of imperfect CSI (ICSI) knowledge at the APs in Fig. \ref{sinr_availability_csi_jt_zf_epa_k4_t4}. With JT and ZF we observe a non-negligible loss of about $6$ dB in the SINR availability when considering $J=1$. This loss however decreases when increasing $J$ to about $1$ dB with $J=16$, showing that, even when the CSI is not perfectly known, distributed MIMO with ZF provides huge gains when compared to centralized MIMO.

\begin{figure}[t!]
\centering
\includegraphics[width=1\hsize]{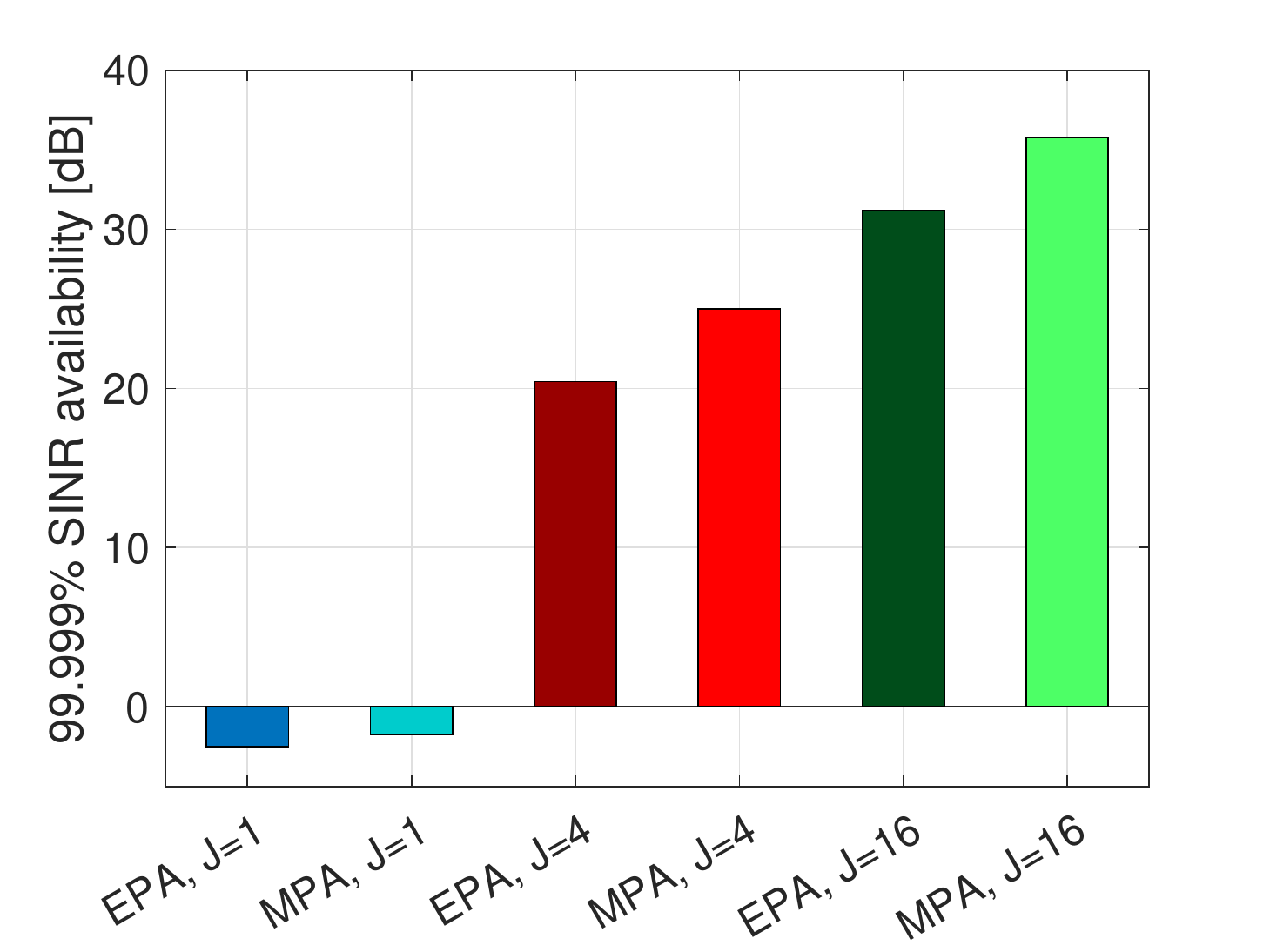}\\
\caption{Impact of MPA on the SINR availability with JT and ZF for different deployments by assuming ICSI and $K=T=4$.}
\label{sinr_availability_pa_jt_zf_icsi_k4_t4}
\end{figure}

In Fig. \ref{sinr_availability_pa_jt_zf_icsi_k4_t4} we analyze, for the same setup of Fig. \ref{sinr_availability_csi_jt_zf_epa_k4_t4}, the gain provided by the MPA algorithm proposed in Sect. \ref{sect_mpa}. Allocating the power in order to maximize the minimum SINR among the active ACs increases fairness in the system, as the APs allocate more power to the ACs experiencing high path-loss and less power to the ACs either close to the APs or creating too much interference in the system. Results show that with ZF and distributed MIMO the gain achieved by MPA in the SINR availability goes up to about $5$ dB when compared to EPA.

\begin{figure}[t!]
\centering
\includegraphics[width=1\hsize]{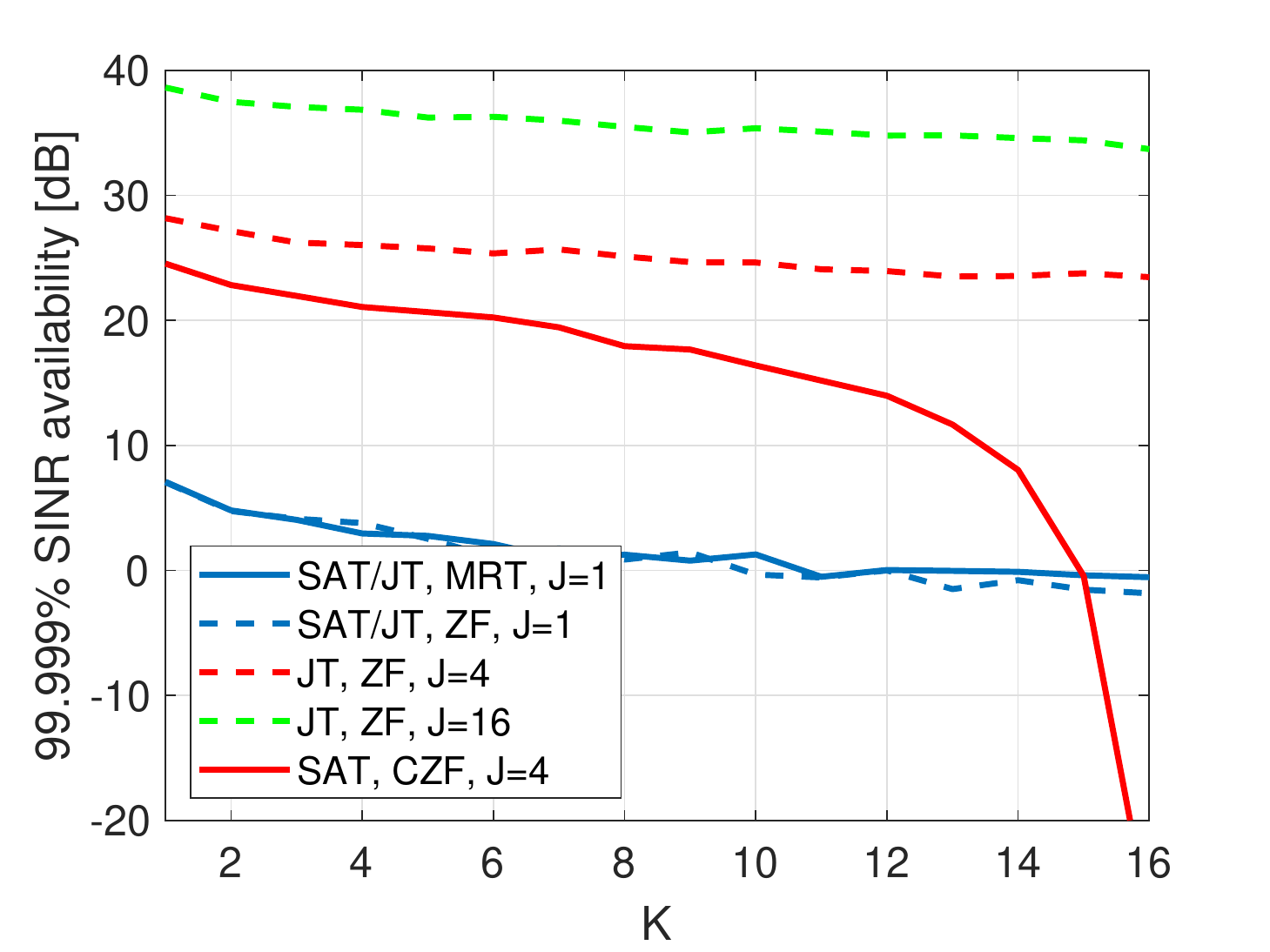}\\
\caption{SINR availability achieved by different configurations for increasing values of $K$ by assuming ICSI, MPA and $T=16$.}
\label{sinr_availability_k_mpa_icsi_t16}
\end{figure}

In Fig. \ref{sinr_availability_k_mpa_icsi_t16} we compare the SINR availability of some of the considered transmission modes and beamformers with MPA and ICSI for different values of $K$: to avoid interference on the channel estimation, we assume for all the configurations $T=16$ in (\ref{eq_impcsi}), which is necessary to avoid any pilot reuse for the maximum value $K=16$ shown here. As expected, the SINR availability decreases when $K$ increases for all the configurations, because less power is allocated to each AC and more interference is present in the system with higher $K$. With centralized MIMO, ZF and MRT provide very similar performance, whereas distributed MIMO with JT and ZF strongly outperforms centralized MIMO by providing a gain of about 20 and 30 dB even with $K=16$ when considering $J=4$ and $J=16$, respectively. Eventually, SAT with $J=4$ and ZF provides very good SINR availability, but only when the number of spatially multiplexed ACs $K$ is sufficiently lower than the number of antennas $M$ at each AP ($K$ up to $14$ in this example with $M=16$). These results confirm that distributed MIMO is an important enabler to increase availability for factory automation. Moreover, for low-to-moderate traffic load, i.e., lower values of $K$, SAT with CZF is probably the best compromise by providing good performance but with a lower cost, as SAT, differently from JT, does not require data sharing among the APs. However, when the traffic load is high, JT is necessary to achieve good SINR values to ensure a reliable transmission.

\begin{figure}[t!]
\centering
\includegraphics[width=1\hsize]{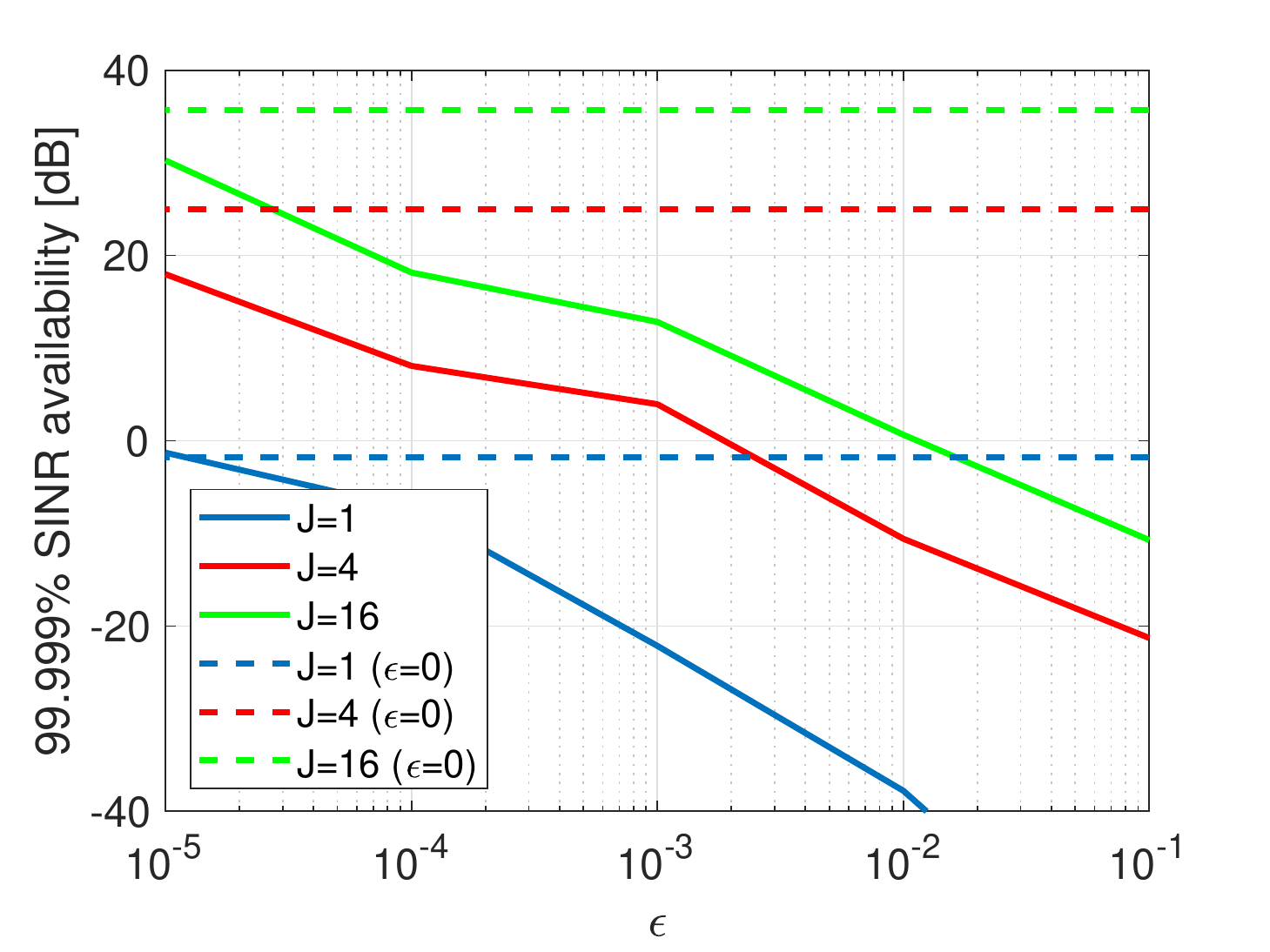}\\
\caption{Impact of the impulsive noise on the SINR availability with JT and ZF for different deployments by assuming MPA and $K=T=4$: dashed lines denote the upper bounds obtained by considering only the thermal noise.}
\label{sinr_availability_impnoise_jt_zf_mpa_icsi_k4_t4}
\end{figure}

Finally, the impact of the impulsive noise is shown in Fig. \ref{sinr_availability_impnoise_jt_zf_mpa_icsi_k4_t4} where, when considering JT with MPA and ICSI (with $T=K=4$), we report the SINR availability for $\Gamma = 30$ dB \cite{bhatti_eusipco09} and different values of the probability $\epsilon$ of the impulsive noise event. The presence of the impulsive noise (without any specific countermeasure) strongly degrades the system performance: for instance, with $J=16$ we observe a loss of about $15$ dB in the SINR availability by assuming just $\epsilon=10^{-4}$, which is related anyhow to a rather rare impulsive noise event. For higher values of $\epsilon$, the performance degradation is even higher. Although these results were obtained with the very basic model in (\ref{eq_impnoise}), they give already an indication that the impulsive noise in industrial environments is potentially a major source of performance degradation, which requires a better understanding, modeling, and a deep study of techniques to mitigate it.

%Finally, in order to understand which kind of rate can be guaranteed in this system, we report in Fig. \ref{throughput_availability_pa_jt_zf_icsi_k4_t4} the achievable throughput corresponding to the SINR availability in the setup of Fig. \ref{sinr_availability_pa_jt_zf_icsi_k4_t4}. As most of the use cases in factory automation are characterized by traffic with small packets, for the spectral efficiency computation we use the model proposed in \cite[(2)]{xu_twc16}, with a codeword length of $512$ bits and a BLER of $10^{-8}$.
%\begin{figure}[t!]
%\centering
%\includegraphics[width=1\hsize]{./Figures/throughput_availability_pa_jt_zf_icsi_k4_t4.eps}\\
%\caption{Throughput availability with JT and ZF for different deployments by assuming ICSI and $K=T=4$.}
%\label{throughput_availability_pa_jt_zf_icsi_k4_t4}
%\end{figure}

\section{Conclusions}

In this paper, we have studied distributed MIMO schemes as a way to improve the SINR availability for factory automation. In order to provide a comprehensive understanding, we have performed extensive system simulations a) with a recent channel model based on measurements done in 5G bands and b) considering the impact of the impulsive noise, that characterizes several types of machines often present in an industrial environment. In this scenario, we have developed a max-min power allocation algorithm specifically improving the SINR availability and compared different deployments, transmission modes, and beamformers. Numerical results have shown that distributed MIMO with ZF and MPA provides a huge gain in the order of 30 dB when compared to centralized MIMO. Moreover, for URLLC communications in general and for factory automation in particular, exactly because we look at extreme requirements like a $99.999\%$ system availability, a good modeling of even rare events is fundamental. In that context, the impulsive noise, even when not appearing that often, can strongly degrade the performance. Future works will include an analysis of distributed MIMO with different AC-AP association rules \cite{buzzi_cml17}, and with more detailed models for the small scale fading and for the impulsive noise.

\section{Acknowledgment}

The work of Stefano Buzzi has been supported by the MIUR program ``Dipartimenti di Eccellenza 2018-2022".

%\IEEEtriggeratref{5} % Command to equalize the two coloumns of the last page
\balance

\bibliographystyle{IEEEtran}
\bibliography{IEEEabrv,full_bibliography}

\end{document}

%% file: macro.tex
\usepackage{amsthm}

\usepackage{enumitem}
\newcounter{problem}
\newcounter{subproblem}[problem]

\usepackage{balance}

\usepackage{hyperref}

\usepackage{mathtools}
\usepackage{color}
\usepackage{xcolor}
\usepackage{colortbl}
\definecolor{purple}{RGB}{139, 0, 139}
\usepackage{manfnt}
\usepackage{cleveref}
\usepackage{url}

\newif\iftodo   % L"a"st \todo-Eintr"age zu ('Baustellen/Unfertiges')
\todotrue
\newif\iftodoshort  % true: Druckt nur den \todo-Marker ohne Kasten aus
\todoshortfalse
\ifCLASSINFOpdf
  \usepackage[pdftex]{graphicx}
   \DeclareGraphicsExtensions{.pdf,.jpeg,.png,.tiff}
\else
   \usepackage[dvips]{graphicx}
   \DeclareGraphicsExtensions{.eps,.ps}
\fi
\usepackage{amssymb,amsmath}
\usepackage[mathscr]{euscript}
\usepackage{bm} 
\usepackage{bbm}
\usepackage{dsfont}

\usepackage{algorithmic}
\usepackage[ruled,vlined,commentsnumbered]{algorithm2e}

\graphicspath{ {./fig/} }
\newcommand{\Rmnum}[1]{\uppercase\expandafter{\romannumeral #1}}
\newcommand{\rmnum}[1]{\lowercase\expandafter{\romannumeral #1}}

\usepackage[utf8]{inputenc}

\usepackage{hyperref}
\usepackage{amsthm}
\usepackage{amssymb,amsmath}
\usepackage[mathscr]{euscript}

\newcommand{\cosl}[1]{}
\newcommand{\resl}[1]{}

\usepackage[draft,footnote,nomargin]{fixme}
\newcommand{\fnql}[1]{}
\newcommand{\fnsv}[1]{}